\documentstyle{article}

\def\beq{\begin{equation}}
\def\eeq{\end{equation}}
\def\pd{\partial}
\def\bea{\begin{eqnarray}}
\def\eea{\end{eqnarray}}

\renewcommand{\thefootnote}{\fnsymbol{footnote}}
\begin{document}
\vspace{0.6cm}
\begin{center}
{\Large{\bf Poisson structures in BRST--antiBRST invariant Lagrangian
 formalism}}\\
\vspace{1cm}

{\large Bodo Geyer}$^{a}$\footnote{e-mail: geyer@itp.uni-leipzig.de},
{\large Petr Lavrov}$^{a, b}$\footnote{e-mail: lavrov@tspu.edu.ru} and
{\large Armen Nersessian $^{a,c,d}$}\footnote{e-mail: nerses@thsun1.jinr.ru}

\vspace{0.5cm}
{\normalsize\it $^{a)}$ Center of Theoretical Studies, Leipzig University,
Augustusplatz 10/11, D-04109 Leipzig, Germany}\\
{\normalsize\it $^{b)}$ Tomsk State Pedagogical University,
634041 Tomsk, Russia}\\
 {\normalsize\it $^{c)}$ Laboratory of Theoretical Physics, Joint Institute
for Nuclear Research, Dubna,  141980 Russia}\\
{\normalsize\it $^{d)}$ Yerevan State University, A. Manoogian St., 3,
Yerevan, 375025, Armenia}
\end{center}
\smallskip
\begin{abstract}
We show that the specific operators $V^a$ appearing in the triplectic 
formalism --- the most general $Sp(2)$ symmetric 
Lagrangian BRST quantization scheme --- 
can be viewed as the anti-Hamiltonian vector fields generated by
a second-rank irreducible $Sp(2)$ tensor. This allows for an explicit 
realization of the ``triplectic algebra'' being constructed from an 
arbitrary Poisson bracket on the space of the fields only, equipped by the
flat Poisson connection.
We show that the whole space of fields and antifields can be 
equipped by an even supersymplectic structure, when this Poisson bracket 
is nondegenerate.
This observation opens the possibility to provide the 
BRST/antiBRST path integral by a well-defined integration measure, 
as well as to establish a direct link between $Sp(2)$ symmetric 
Lagrangian and Hamiltonian BRST quantization schemes.
\end{abstract}
\bigskip
\setcounter{page}1
\renewcommand{\thefootnote}{\arabic{footnote}}
\setcounter{footnote}0
\setcounter{equation}0
{\bf  Introduction.}
The Batalin-Vilkovisky (BV) formalism \cite{BV} of Lagrangian 
quantization of general gauge theories, since its introduction, 
attracts permanent interest due to its covariance, universality 
and mathematical elegance. Now, its area of physical applications is 
much wider than offered in the initial prescription. The BV-formalism 
is outstanding also from the formal mathematical point of view
because it is formulated in terms of seemingly exotic objects: the
antibracket (odd Poisson bracket) and the (related) second-order operator 
$\Delta$.

The study of the formal geometrical structure of the BV formalism,
performed during the last ten years, allowed to introduce its 
interpretation in terms of more traditional mathematical objects 
\cite{bvgeom,Kh}, as well as to find the unusual behaviour of the 
antibracket with respect to integration theory \cite{K}.
On the other hand, there exists an extended, 
$Sp(2)$ symmetric (BRST/antiBRST invariant) version of the
BV formalism \cite{BLT}, and its geometrized version  
known as ``triplectic formalism'' \cite{BM,BMS} (see also
\cite{ND,GGL}). In addition, there exists also a $Sp(2)$ invariant
extension whose structure is characterized by the superalgebra
$osp(1,2)$ \cite{GLM}.

In the BV formalism the original set
 of ``physical'' fields $\phi^A$ (including ghosts, antighosts, 
Lagrangian multipliers etc.) is doubled  by  the 
``antifields'' $\phi^*_A$ with opposite grading $\epsilon(\phi^*_A)=
\epsilon(\phi^A)+1$. On this set of fields and antifields
the {\it nondegenerate} antibracket and the corresponding $\Delta-$operator 
are defined.
In the $Sp(2)$ symmetric versions of the BV-formalism the initial set 
of fields $\phi^A$ is extended by additional auxiliary fields 
$\bar\phi_A$ as well as the $Sp(2)$ doublets $\phi^*_{Aa}$ and 
$\pi^{Aa}$ ($a=1,2$) of antifields with gradings 
$\epsilon(\bar\phi_A)=\epsilon(\phi^A)$, 
$\epsilon(\phi^*_{Aa})=\epsilon(\pi^{Aa})=\epsilon(\phi^A)+1$;
the antifields $\pi^{Aa}$ are used in the formulation of gauge 
fixing conditions.
As usual, the $Sp(2)$ indices are raised and lowered by
the help of $\varepsilon^{ab}$ and $\varepsilon_{ab}$:
$\varepsilon^{ac}\varepsilon_{cb}= \delta^a_b,\;
 \varepsilon_{ab}=-\varepsilon_{ba},\;\varepsilon_{12}=1$.

On this extended space of fields, ${\cal M}$, the basic ingredients of 
the $Sp(2)$ 
symmetric  Lagrangian quantization schemes are formulated by pairs of 
$\Delta$-operators, $\Delta^a$, and odd vector fields, $V^a$, 
satisfying the following consistency conditions:
\begin{eqnarray}
&\Delta^{\{a}\Delta^{b\}} =0~,\quad V^{\{a}V^{b\}} =0~,&
\quad\quad\epsilon(V^a)=\epsilon(\Delta^a)=1~, \label{dd} \\
%
%
&\Delta^{\{a} V^{b\}} + V^{\{a}\Delta^{b\}} = 0 .&
\label{vdw}
\end{eqnarray}
Here and in the following the curly bracket denotes symmetrization with
respect to the indices $a$ and $b$.
These operators are formulated in terms of $\phi^A,
\bar\phi_A, \phi^*_{Aa}$ fields in the original scheme \cite{BLT},
and on the whole set $z = \{\phi^A, \bar\phi_A, \phi^*_{Aa}, \pi^{Aa}\}$ 
in the case of triplectic formalism \cite{BM,BMS}.
An essential difference between the $Sp(2)$ symmetric quantization schemes 
and the   standard BV formalism is not only the doubling of 
$\Delta-$operators and corresponding antibrackets, but also
{\it the appearance of the vector fields $V^a$} and 
{\it the degeneracy of the antibrackets} generated by 
the $\Delta^a$--operators as obstruction of the Leibniz rule,
\beq
(-1)^{\epsilon(f)}(f,g)^a = \Delta^a(fg) - (\Delta^a f) g -
 (-1)^{\epsilon(f)}f (\Delta^a g)~. 
\label{liebdelta} 
\eeq
These antibrackets  obey the  conditions \cite{BLT}:
\begin{eqnarray}
 & &(f,g)^a=-(-1)^{(\epsilon(f)+1)(\epsilon(g)+1)}(g,f)^{a}~,
\label{as}\\
 & &(-1)^{(\epsilon(f)+1)(\epsilon(h)+1)}(f,(g,h)^{\{a})^{b\}} + ~
 {\mbox{\rm cycl. perm.}}~ = 0~, 
\label{bj}\\
 & &\Delta^{\{a}(f,g)^{b\}} =(\Delta^{\{a} f,g)^{b\}}
 +(-1)^{\epsilon(f)+1} (f,\Delta^{\{a} g)^{b\}}~,    
\label{ad}\\
& &V^{\{a}(f,g)^{b\}} = (V^{\{a}f,g)^{b\}}+(-1)^{\epsilon (f)+1} 
(f, V^{\{a}g)^{b\}}~. \label{avw}
\end{eqnarray}
In the triplectic quantization schemes \cite{BM,BMS}
much stronger compatibility conditions are
fulfilled, namely, instead of (\ref{vdw}) and (\ref{avw}), 
one requires, respectively
\beq
\Delta^{a} V^{b} + V^{a}\Delta^{b} = 0~,\quad
 V^a(f,g)^b = (V^af,g)^b  +(-1)^{\epsilon (f)+1} (f, V^a g)^b~.
\label{av}
\eeq

The partition function in the triplectic formalism
is defined by the expression
\begin{equation}
Z=\int [dv][d\lambda]{\rm e}^{\frac{i}{\hbar}\left[W(z) +
X(z,\lambda)\right]}
{}~,\label{z}
\end{equation}
where $[dv]$ stands for the volume element,
$W(z)$ is viewed as the quantum action of the theory, 
and $X(z,\lambda)$ is considered as gauge fixing term.
$\lambda$ are additional auxiliary variables which simply become
Lagrangian multipliers of the gauge constraints when $X$ depends on them
only linearly.
The partition function (\ref{z}) is gauge independent if the following
``quantum master equations'' hold \cite{BMS}:
\beq
(\Delta^a +\frac{i}{\hbar} V^a){\rm e}^{\frac{i}{\hbar} W} =
0~\;\;
\Leftrightarrow\;\;
-i \hbar\Delta^a W+ V^aW+ \frac 12(W,W)^a=0~,
\eeq
as well as  a similar equation for the gauge fixing functional
$(\Delta^a-\frac{i}{\hbar}V^a + \ldots)
{\rm e}^{\frac{i}{\hbar} X} = 0$ (the dots indicate the extra 
terms that are required due to the variation of the variables $\lambda$).
The (anti)BRST transformations are generated by the operators $\delta^a$,
\beq
\label{brstanti}
 \delta^a=(W -X,\,\cdot\,)^a+2V^a +i\hbar\Delta^a~.
\eeq

An important feature of the triplectic algebra which was observed in
 Ref.~\cite{BMS} is the existence of some Lagrangian submanifold 
${\cal M}_0$ on which the following $Sp(2)$ invariant, 
 even Poisson bracket can be defined, viz.
\beq
\{u,w\}_0 \equiv \epsilon_{ab}(u, V^aw)^b ~.
\label{even}
\eeq
This submanifold ${\cal M}_0$ is specified by the  requirement that
all the functions   $u(x), v(x), w(x)$ on ${\cal M}_0$ should obey the 
following conditions:
\beq
(u,v)^a = 0~,\quad  (u,V^{\{a} v)^{b\}},w)^c=0~.
\label{lagr}\eeq
In that case it can be checked, by the use of 
  Eqs.~(\ref{as}), (\ref{bj}) and (\ref{avw}), 
that the operation (\ref{even})  defines an {\it even Poisson bracket} 
on ${\cal M}_0$ obeying the following relations:
\beq
\begin{array}{c}
\{u,v\}_0  = -(-1)^{\epsilon(u)\epsilon(v)} \{v,u\}_0~,
\\\\
\{u,vw\}_0 = \{u,v\}_0w + (-1)^{\epsilon(u)\epsilon(v)}v\{u,w\}_0 ~,
\\\\
(-1)^{\epsilon(u)\epsilon(w)}\{\{u,v\}_0,w\}_0  +
{\rm cycl.}\;{\rm perm.}  = 0 ~.
\end{array}
\label{pb}
\eeq
In the canonical versions \cite{BLT,BM} 
of the $Sp(2)$ covariant 
Lagrangian quantization schemes this construction yields a canonical 
Poisson bracket on the fields $\phi^A$ and $\bar\phi_A$, namely,  
$\{\phi^A,\bar\phi_B\}_0=\delta^A_B$.

Some efforts were performed to get the possible restrictions on this 
Poisson bracket and to clarify its role in the triplectic formalism.
First of all, in Ref.~\cite{GS}, it was established that it is possible 
to construct the triplectic algebra when ${\cal M}_0$ is a special type 
of K\"ahler manifold; then, the triplectic algebra with 
{\it nondegenerate antibrackets} was constructed on the 
space of differential forms (on the external algebra) of arbitrary 
K\"ahler manifolds \cite{BN}.
In the latter case the vector fields $V^a$ may be interpreted as the 
holomorphic and antiholomorphic external differentials, while
the whole supermanifold was equipped with an {\it even K\"ahler structure}
(see also \cite{KN}).

In this note we establish two properties of the triplectic formalism, viz.
\begin{itemize}
\item 
the explicit realization of the triplectic algebra over 
an {\em arbitrary flat  Poisson manifold} (i.e. the Poisson manifold equipped
 with flat connection respecting Poisson bracket) is given;
\item the triplectic formalism is equipped with an even symplectic structure.
\end{itemize}
The suggested structures seem to be important 
\begin{itemize}
\item 
for the construction of an integration measure in the triplectic formalism, 
and 
\item 
for  establishing a direct link between the Hamiltonian
\cite{BLTH} and the Lagrangian \cite{BLT,BM,BMS} 
BRST-antiBRST invariant quantization schemes; 
\end{itemize}
these are problems which, despite of some efforts on that subject \cite{TGG},
are still to be solved. 

In our consideration we assume, for the simplicity, that ${\cal M}_0$
is a Poisson manifold, though in the BRST quantization formalisms it
necessarily contains also Grassmannian degrees of freedom (e.g.,
ghosts/antighosts).
Of course, the transport of the proposed constructions
to supermanifolds is straightforward. \\
 \\
{\bf  Poisson bracket and $V^a$ fields}. Let us
consider a Poisson manifold 
 $({\cal M}_0,  \{\,\cdot\,,\,\cdot\,\}_0)$ 
where the (even) Poisson bracket obeying 
Eqs.~(\ref{pb}) is given by 
\beq
\{f(x), g(x)\}_0=\frac{\partial f(x)}{\partial x^i}
\omega^{ij}(x)\frac{\partial g(x)}{\partial x^j}~,\quad\quad
\omega^{ij}=-\omega^{ji}~,
\quad \omega^{ij}_{\phantom{ij},l}\omega^{lk}+ \;
{\rm cycl.~perm.}\;=0~.
\label{pbx}
\eeq
Then we consider the superspace ${\cal M}$ parametrized by the
coordinates  $x^i$ and $\theta_{ia}$, 
($\epsilon(\theta_{ia})=\epsilon(x^i)+1=1$), 
where $\theta_{ia}$ (``antifields'') are 
 transformed as $\partial/\partial x^i$ under reparametrizations of 
the base manifold ${\cal M}_0$ with coordinates $x^i$ (``fields'').
In Darboux coordinates one has
$x^i = (\phi^A, {\bar\phi}_A), 
\theta_{ia} = (\phi_{Aa}^*, \varepsilon_{ab}\pi^{Ab})$.

On ${\cal M}$ we introduce the triplet of functions $S_{ab}$
defining an irreducible second rank $Sp(2)$ tensor,
\beq
 S_{ab}=\hbox{\large$\frac{1}{6}$}
\theta_{ia}\omega^{ij}(x)\theta_{jb}~, 
\qquad 
 S_{ab}=S_{ba}~,
\eeq
and  the following  pairs of $\Delta$-operators and  antibrackets
transforming as  $Sp(2)$ doublets,
\beq
\Delta^a=\nabla_i\frac{\pd_l}{\pd\theta_{ia}} +
\frac{\pd\rho}{\pd x^i}\frac{\pd_l}{\pd\theta_{ia}}~,
\qquad
(f,g)^a={\nabla_i f}
\frac{\partial_l g}{\partial \theta_{ia}}
+(-1)^{\epsilon(f)}\frac{\partial_l f}{\partial \theta_{ia}}
\nabla_i g~,
\label{abf}
\eeq
where $\nabla_i$ 
corresponds to the {\it flat symmetric connection} $\Gamma^k_{ij}(x)$
which respects  the Poisson bracket (\ref{pbx}) \cite{fm}:
\beq
\nabla_i\equiv\frac{\partial}{\partial x^i}+
\Gamma^k_{ij}(x)\theta_{ka}\frac{\partial}{\partial\theta_{ja}}
:\qquad 
\pd_k\omega^{ij}+\Gamma^i_{kl}\omega^{lj}+\omega^{il}\Gamma^j_{lk} =0,\quad
 \Gamma^k_{ij}=\Gamma^k_{ji},\quad
[\nabla_i,\nabla_j]=0~,
\eeq
while $\rho$  obeys the condition 
$\pd_i\omega^{ij} = \omega^{ji}{\pd_i\rho}$ (for nondegenerate Poisson
 brackets it yields the expression 
 $\rho=-\frac 12\,\log\,{\det}\;\omega^{ij}$).  
When the Poisson bracket is nondegenerate,
the   flat connection exists,  at least  locally,  due to  the 
Darboux theorem, since in canonical  coordinates one can choose the trivial 
connection which is obviously flat \cite{btn,kf}.
Let us now introduce the (nilpotent) vector fields $V_a$ being generated 
by the 
antibrackets through the irreducible second rank $Sp(2)$ tensor $S_{ab}$
according to:
\beq
V_a=(S_{ab}, \,\cdot\,)^b=
-\hbox{\large$\frac 12$} \theta_{ia}\omega^{ij}\nabla_j~.
\label{vs}
\eeq  
These  fields can also be represented in the form (no summation)
\beq
V_a=3(S_{11}+S_{22},\,\cdot\,)^a=6(-1)^a(S_{12},\,\cdot\,)_a~,
\eeq
i.e., these vector fields are anti-Hamiltonian with respect
to both antibrackets, and they obey the conditions (\ref{av}).
Hence they form, together with the antibrackets $(\,\cdot\,, \,\cdot\,)^a$ 
and the $\Delta^a-$ operators (\ref{abf}), 
the triplectic algebra (\ref{dd}),
(\ref{liebdelta}) -- (\ref{ad}) and (\ref{av}).

{\it According to that procedure we constructed an explicit realization 
of the triplectic algebra on $\cal M$ with an arbitrary flat  
 Poisson manifold ${\cal M}_0$.} \\
\\
{\bf $U^a$ fields.} Let us now introduce  particularly important 
realizations of anti-Hamiltonian 
vector fields  generated  by a $Sp(2)$ scalar function 
$S_0$ according to
\beq
U^a = (S_0, \,\cdot\,)^a~: 
 \qquad 
S_0=\hbox{\large$\frac 12$}
 \varepsilon^{ab}\theta_{ia}g^{ij}(x)\theta_{jb}~, 
 \quad\epsilon(S_0) = 0 ~,\quad 
 g^{ij}=g^{ji}~,
\label{VHam}
\eeq
with some (graded) symmetric tensor $g^{ij}$.
Substituting these expressions into (\ref{even}) yields a Poisson bracket
which vanishes identically on ${\cal M}_0$ (even without requiring their 
nilpotency).

Such vector fields $U^a$ appear within the modified triplectic quantization 
\cite{GGL}. In that case they are required to obey also the conditions
\beq
U^{\{a}U^{b\}}=0~,\quad V^{\{a}U^{b\}}+ U^{\{a}V^{b\}} =0~,\quad
 \Delta^{\{a}U^{b\}}+ U^{\{a}\Delta^{b\}} =0~,
\eeq
which are equivalent to the following ones,
\beq
 (S_0, S_0)^a=0~,\quad  V^a S_0=0~,\quad \Delta^aS_0=0~.
\label{s0}
\eeq
Now, it is convenient to introduce $I^i_j(x)=g_{jk}\omega^{ik}$,
where $g^{ij}=\omega^{ik}g_{kl}\omega^{lj}$, and to interpret
$g_{ij}$ as a ``metric tensor'' on ${\cal M}_0$.
In that case Eqs.~(\ref{s0}) imply the following conditions
for $I^i_j$:
\beq
I^k_{i;j}-I^k_{j;i}=0,\quad I^n_jI^i_{n;k}-I^n_kI^i_{n;j}=0
\eeq 
where the semicolon ``$\ ;\ $'' denotes the covariant derivative.
Hence, the Nijenhuis tensor $N^i_{jk}(I)$ of $I^i_j$ vanishes:
\beq
N^i_{jk}(I)\equiv
I^n_jI^i_{n;k}-I^n_kI^i_{n;j}-I^i_n(I^n_{j;k}-I^n_{k;j})=0~.
\eeq 
The latter  equations can be resolved
if one requires ${\hat I}$ to be an almost complex structure, 
${\hat I}^2=-{{\rm id}}$, i.e.,  when  
${\cal M}_0$ is a  K\"ahler manifold.

{\it So, on K\"ahler manifolds one can consistently formulate not only 
the triplectic algebra, but also the modified triplectic algebra.} \\
\\
{\bf Even Symplectic Structure}. When the Poisson bracket (\ref{pbx}) 
is non-degenerate, the superspace ${\cal M}$
can be equipped with both the {\it even symplectic structure} and the
corresponding {\it non-degenerate Poisson bracket},
\bea
\Omega=dz^\mu\Omega_{\mu\nu}dz^\nu
&=&\omega_{ij}dx^i\wedge dx^j+\alpha^{-1}
\omega^{ij}D\theta_{ia}\wedge D\theta^a_j~,
\nonumber\\
 \{f(z), g(z)\}
&=&(\nabla_i f)\omega^{ij}( \nabla_j g)+
\alpha\frac{\partial_r f}{\partial \theta^a_{i}}
\omega_{ij}\frac{\partial_l g}{\partial \theta_{ja}}~,
\label{ss}
\eea
where $\alpha$ is an arbitrary constant, and 
$D\theta_{ia}=d\theta_{ia}-\Gamma^k_{ij}\theta_{ka}dx^j$.
 The  proposed two-form $\Omega$ is closed
(and, hence, the Jacobi identity for the Poisson bracket holds)
 due to the flatness of the symplectic connection 
and the equality $\omega^{ij}D\theta_{ia}\wedge D\theta^a_j=
d(\theta_{ia}\omega^{ij} D\theta^a_j)$.

Using the even Poisson bracket as introduced above we can define the 
 analogue of the Liouville measure on symplectic (super)surfaces
$\Upsilon$ (and, as a consequence, on the whole superspace
 ${\cal M}$)  defined by the help 
of the equations $f^\mu (x,\theta_a)=0$ (cf., Ref.~\cite{KGS}) as follows:
\beq
{\cal D}_{\Upsilon}(x,\theta|f^\mu)=
{\sqrt{{\rm Ber}\{f^\mu,f^\nu\}
}}\,\delta(f^{\mu})\,{\cal D}_0~,\quad
{\cal D}_0={\sqrt{
{\rm Ber}\;\Omega_{\mu\nu}}}=\alpha\left({{\det\;\omega_{ij}}}\right)^{3/2}~.
\label{rho}
\eeq
These measures are invariant under supercanonical transformations 
of the Poisson bracket (\ref{ss}), as well as under smooth deformations 
of the surface $\Upsilon$ \cite{KGS}.

Note, that the antibrackets (\ref{abf})
may be connected with the proposed even 
symplectic structure as follows:
$\Omega_a=\alpha {\cal L}_a\Omega$,
where ${\cal L}_a$ denotes the Lie derivative along the vector 
field $V^a$,
while the closed two-forms  $\Omega_a\equiv dx^i\wedge D\theta_{ia}=
dx^i\wedge d\theta_{ia}$,
are pseudo-inverse to the antibrackets (\ref{abf}).
 The (super)group of transformations preserving
simultaneously the Poisson bracket  $\{\,\cdot\,,\,\cdot\,\}$ and the 
antibrackets $(\,\cdot\,,\,\cdot\,)^a$ is infinite-dimensional,
 due to the degeneracy of the antibrackets (while
 the group of transformations preserving simultaneously even
 and odd symplectic structures is necessarily finite-dimensional 
 \cite{Kh}).
At this stage it appears as a natural desire to establish, by the use of 
the symplectic structure (\ref{ss}), a direct link between the 
$Sp(2)$ invariant Lagrangian and Hamiltonian BRST formalisms.
Furthermore, one may tend to use the measure (\ref{rho}) in the path 
integral of the triplectic formalism.
\\
\\
{\bf Discussion and summary.}
In this note we made some observations on the formal structure 
of the triplectic formalism in its ``strong'' and modified version.

First, we established that the vector fields $V^a$ appearing 
in the triplectic formalism can be represented as anti-Hamiltonian 
vector fields generated by the Poisson bracket through an
irreducible second rank $Sp(2)$ tensor.
This allowed us to give a realization of the triplectic algebra, in an
arbitrary coordinate system, by using a {\it flat Poisson  connection}.
Note, that the flat connection respecting Poisson bracket 
was used by I. A. Batalin and I. V. Tyutin for the formulation of 
coordinate-free scheme of deformation quantization \cite{btn}.
The symmetric connections which respect the symplectic structure
are  widely  known under a nickname ``Fedosov connections'' \cite{fm}
due to  Fedosov's works on globally defined
   deformation quantization \cite{F}. These connections  were  
recently found to be the natural objects in the Hamiltonian BRST 
quantization as well \cite{BGL}.
We have also shown that, requiring additionally the base Poisson 
manifold to possess a K\"ahler structure, allows to equip the triplectic
algebra with additional vector fields $U^a$, appearing in modified 
triplectic quantization. These operators are responsible for the
gauge fixing with a separation into ``physical variables'' and their
``momenta''.

Second, we have found that the triplectic algebra can be equipped by an
{\it even symplectic structure}.  
This symplectic structure could be used 
for the construction of an integration measure in the triplectic formalism, 
as well as for establishing a direct link between Lagrangian and 
Hamiltonian BRST quantization schemes.
Note, that the existence of the even symplectic structure 
looks to be a natural one in the the superfield 
 formulation of the BRST quantization \cite{superH,LM}.
The symplectic structure $\Omega$ and the odd 
two-forms  $\Omega_a$ could be extracted
from the closed superfield two-form
$\omega(\Phi)_{ij}d\Phi^i\wedge d\Phi^j$, where 
$\Phi^i=x^i+\omega^{ij}(\eta^a\theta_{ja}+\eta^1\eta^2y_j)$,
$\epsilon(\eta^a)=1$       
and $y_i$ are auxiliary fields denoting, in the flat case, 
sources and Lagrangian multipliers $y_i=(\lambda^A, J_A)$ \cite{LM}.

In our opinion, it seems to be mysterious, that just the BRST/antiBRST 
invariant scheme includes, in its most symmetric (triplectic) formulation, 
all the ingredients appearing in various quantization formalisms. 
It might be assumed, that within that framework it would be possible
 to develop a complete (or universal) quantization scheme containing, 
as special limits, all existing methods,
 both BRST and non-BRST, Lagrangian and Hamiltonian ones.
\vspace{0.5cm}

\noindent
{\sc Acknowledgements:}~The authors thank A. Karabegov for several 
discussions on Fedosov geometry, and I. V. Tyutin for  criticism and useful 
remarks.
P.L. and A.N. acknowledge the hospitality of NTZ at the 
Center of Advanced Study of Leipzig University and financial
  support by the Saxonian
 Ministry of Fine Arts and Sciences, which made this collaboration possible.
 The  work of P.L. was also supported under the projects  RFBR  99-02-16617, 
 INTAS   99-0590,  RFBR-DFG  99-02-04022 and the
Fundamental Sciences Grant E00-3.3-461 of the Russian Ministry of Education.
 The work of A.N. was also supported under the INTAS project 00-A1-0262.


\begin{thebibliography}{99}

\bibitem{BV}
 I.A. Batalin and G.A. Vilkovisky, Phys. Lett. {\bf 102B}
 (1981) 27; Phys. Rev. {\bf D28} (1983) 2567 [E: {\bf D30} (1984) 508];
 Nucl. Phys. {\bf B234} (1984) 106.

\bibitem{bvgeom}
 E. Witten, Mod.\ Phys.\ Lett.\ A {\bf 5} (1990) 487;\\
 O.~M.~Khudaverdian and A.~P.~Nersessian,
 Mod.\ Phys.\ Lett.\ A {\bf 8} (1993) 2377;
 J.\ Math.\ Phys.\ {\bf 37} (1996) 3713.
 \newline
 A. Schwarz, Commun. Math. Phys. {\bf 155} (1993) 249; {\it ibid.}
 {\bf 158} (1993) 373. \newline
 I.A. Batalin and I.V. Tyutin, Int. J. Mod. Phys. {\bf A8} (1993)
 2333; Mod. Phys. Lett. {\bf A8} (1993) 3673; 
{\it ibid.} {\bf A9} (1994) 1707.
 \newline H. Hata and B. Zwiebach, Ann. Phys.(N.Y.) {\bf 322} (1994) 131.
 \newline J. Alfaro and P.H. Damgaard, Phys. Lett. {\bf B334} (1994) 369.
\bibitem{Kh}
 O.~M.~Khudaverdian, J.~Math.~Phys. {\bf 32} (1991), 1934
 (preprint UGVA-DPT 1989 / 05 -- 613).

\bibitem{K} 
 O.~M.~Khudaverdian and R.~L.~Mkrtchian, Lett.\ Math.\ Phys.\ {\bf 18}, 
 229 (1989),\\
 O.~M.~Khudaverdian,
 Commun.\ Math.\ Phys.\ {\bf 198}, 591 (1998); 
``Semidensities on odd symplectic manifolds'', [math/DG/0012256] 


\bibitem{BLT} 
 I.A. Batalin, P.M. Lavrov and I.V. Tyutin, J. Math. Phys. 
 {\bf 31} (1990) 1487; {\it ibid.} {\bf 32} (1990) 532; 
 {\it ibid.} {\bf 32} (1990) 2513.

\bibitem{BM}
 I.A.~Batalin and R.~Marnelius, Phys.\ Lett.\ B {\bf 350} (1995) 44.

\bibitem{BMS}
 I.~A.~Batalin, R.~Marnelius and A.~M.~Semikhatov, Nucl.~Phys.
 {\bf B446} (1995), 249.

\bibitem{ND}
 M.~Henneaux, Phys.\ Lett.\ B {\bf 282} (1992) 372.\\ 
 P.~Gregoire and M.~Henneaux, Phys.\ Lett.\ B {\bf 277} (1992) 459;
Commun.\ Math.\ Phys.\ {\bf 157} (1993) 279.\\
 A.~Nersessian and P.~H.~Damgaard, Phys. Lett. 
 {\bf B355} (1995) 150.\\
 I.A.~Batalin and R.~Marnelius, Nucl.\ Phys.\ B {\bf 465} (1996) 521.\\
 M.~A.~Grigoriev and A.~M.~Semikhatov, Phys.\ Lett.\ B {\bf 417} (1998) 259.\\
 M.~A.~Grigoriev, Phys.\ Lett.\ B {\bf 458} (1999) 499.
\bibitem{GGL}
 B.~Geyer, D.M.~Gitman and P.M.~Lavrov,
 {Mod. Phys. Lett.} {\bf A14} (1999) 661.

\bibitem{GLM}
 B.~Geyer, P.M.~Lavrov and D.~M\"ulsch,
 J. Math. Phys. {\bf40} (1999) 674; {\it ibid.} {\bf 40} (1999) 6189;\\
 B.~Geyer and D. M\"ulsch, {J. Math. Phys.} {\bf 41} (2000) 7304.

\bibitem{GS}M.~A.~Grigoriev and A.~M.~Semikhatov,
 Theor.\ Math.\ Phys.\ {\bf 124} (2000) 1157.

\bibitem{BN}S.~Bellucci and A.~Nersessian, ``K\"ahler geometry and SUSY
 mechanics'', [hep-th/0103005].

\bibitem{KN}A.~P.~Nersessian, Theor.~Math.~Phys. {\bf 96} (1993), 866;
 JETP Lett. {\bf 58} (1993) 66;
 Lecture Notes in Physics {\bf 524} (1997) 90; [hep-th/9811110].
 
\bibitem{BLTH}
 I.~A.~Batalin, P.~M.~Lavrov and I.~V.~Tyutin,
 J.\ Math.\ Phys.{\bf 31} (1990) 6;
 {\it ibid.}\ {\bf 31} (1990) 2708.\\
 P.~Gregoire and M.~Henneaux,
 Phys.\ Lett.\ B {\bf 277} (1992) 459;
 J.\ Phys.\ A {\bf 26} (1993) 6073; 
 Commun.\ Math.\ Phys.\ {\bf 157} (1993) 279.

\bibitem{TGG}
 G.~V.~Grigorian, R.~P.~Grigorian and I.~V.~Tyutin,
 Nucl.\ Phys.\ B {\bf 379} (1992) 304.\\
 F.~De Jonghe, Phys.\ Lett.\ B {\bf 316} (1993) 503.\\
 I.~V.~Tyutin and S.~S.~Shahverdiyev,
 Theor.\ Math.\ Phys.\ {\bf 110} (1997) 109.
 

\bibitem{fm}I.~Gelfand, V.~Retakh and M.~Shubin, 
Adv. Math. {\bf 136} (1998) 104, [dg-ga/9707024]
\bibitem{btn}I.~A.~Batalin and  I.~V.~Tyutin,
  Nucl.\ Phys.\ B {\bf 345} (1990) 645.
\bibitem{kf}B.~V.~Fedosov and A.~V.~Karabegov, private communication.
\bibitem{KGS}
 O.~M.~Khudaverdian, A.~S.~Schwarz and Y.~S.~Tyupkin,
 Lett.\ Math.\ Phys.\ {\bf 5}, 517 (1981).
\bibitem{superH}
 I.~A.~Batalin, K.~Bering and P.~H.~Damgaard,
 Nucl.\ Phys.\ B {\bf 515} (1998) 455;
 Phys.\ Lett.\ B {\bf 446} (1999) 175.\\
 M.~A.~Grigoriev and P.~H.~Damgaard,
 Phys.\ Lett.\ B {\bf 474} (2000) 323.

\bibitem{F} B.~V.~Fedosov, J. Diff. Geom. {\bf 40} (1994) 213.

\bibitem{BGL}
 M.~A.~Grigorev and S.~L.~Lyakhovich,
 ``Fedosov deformation quantization as a BRST theory,''
 [hep-th/0003114].\\
 I.~A.~Batalin, M.~A.~Grigoriev and S.~L.~Lyakhovich,
 ``Star product for second class constraint systems from a BRST theory,''
 [hep-th/0101089].
 

\bibitem{LM}
 P.~M.~Lavrov, Phys.\ Lett.\ B {\bf 366} (1996) 160;
 Theor.\ Math.\ Phys.\ {\bf 107} (1996) 602.\\
 P.~M.~Lavrov and P.~Yu.~Moshin,
 ``Superfield Lagrangian quantization with extended BRST symmetry,''
Phys. Lett. {\bf B} (in press) [hep-th/0102106].
\end{thebibliography}
 \end{document}